\documentclass[usenatbib,useAMS]{mn2e}
\usepackage{graphicx}

\title{Pulsar current revisited}
\author[N. Vrane\v{s}evi\'{c} and D. B. Melrose]
{N. Vrane\v{s}evi\'{c}$^{1,2}$ and D. B. Melrose$^1$\\
$^1$School of Physics, University of Sydney, Sydney, NSW 2006, Australia\\
$^2$ATNF, CSIRO, Epping, Australia}
\date{%Accepted \hspace{2cm}
          --- Received in original form}
\pubyear{2010}

\begin{document}

\maketitle
\begin{abstract}
The pulsar current, in the $P$--$ {\dot P}$ plane where $P$ is the pulsar period and ${\dot P}$ is the period derivative, is used as a supposedly ``model free'' way to estimate the pulsar birthrate from statistical data on pulsars. We reconsider the derivation of the kinetic equation on which this is based, and argue that the interpretation of the pulsar current is strongly model dependent, being sensitive to the form of the assumed evolution law for pulsars.  We discuss the case where the trajectory of a pulsar is assumed to be of the form ${\dot P}=KP^{2-n}$ with $K$ and $n$ constant, and show that (except for $n=2$) one needs to introduce a pseudo source term in order to infer the birthrate from the pulsar current. We illustrate the effect of this pseudo source term using pulsar data to estimate the birthrate for different choices of $n$. We define and discuss an alternative ``potential'' class of evolution laws for which this complication is avoided due to the pseudo source term being identically zero.
\end{abstract}

\section{Introduction}

\begin{figure}
\begin{center}
%\vspace{100pt}
\includegraphics[angle=0, width=0.45\textwidth, height=0.3\textheight]{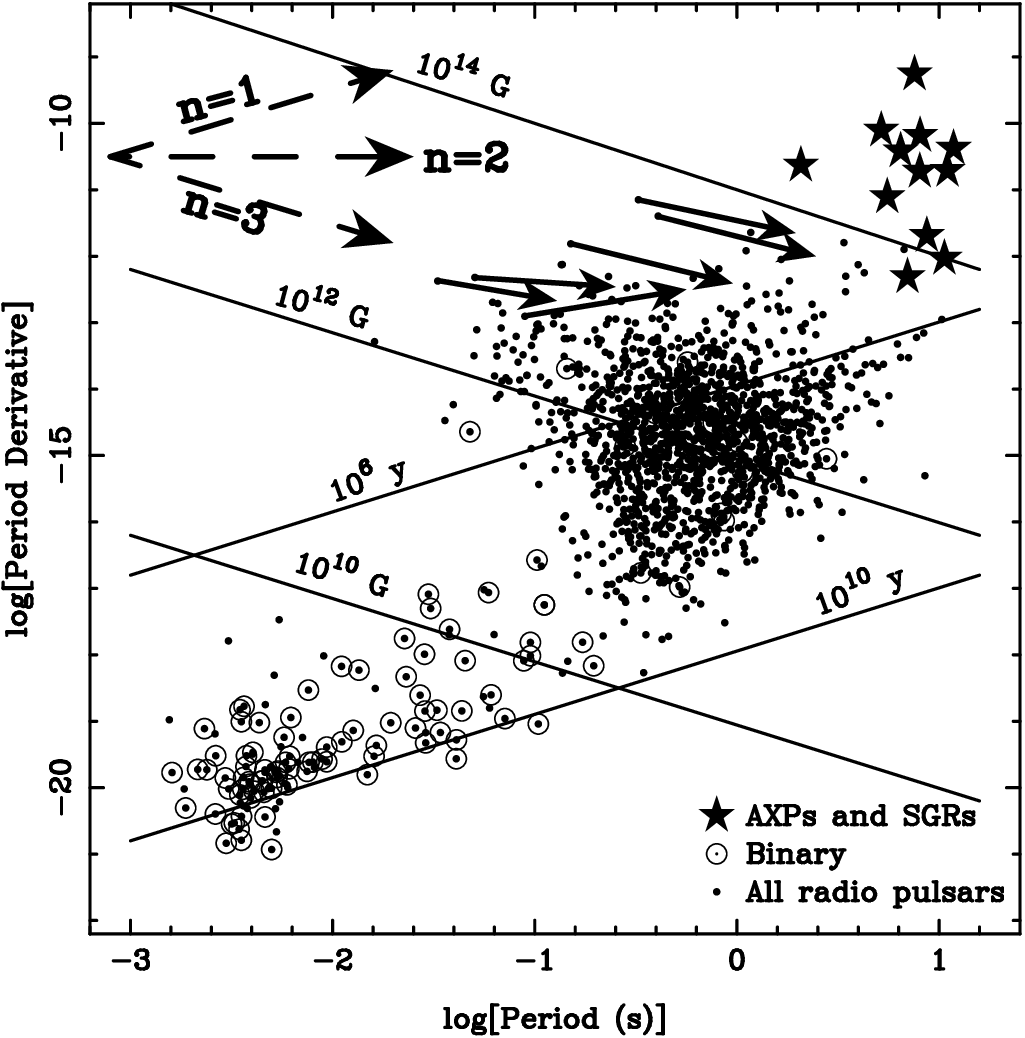}
\caption{A $P$--${\dot P}$ diagram showing the location of different pulsars populations: normal pulsars, millisecond (together with binary) pulsars and AXPs (anomalous X-ray pulsars) \& SGRs (soft $\gamma$-ray repeaters). Dots correspond to all radio pulsars taken from the ATNF pulsar catalogue. Binary pulsars are marked with circles, AXPs \& SGRs are marked with stars. Dashed arrows are showing direction of pulsars motion on the diagram with different braking index values. Solid arrows indicate the current spin evolution of six pulsars for which the braking index has been measured (see Table~\ref{tab1}). Lines of constant characteristic age and surface magnetic dipole field strengths are shown.\label{fig1}}
\end{center}
\end{figure}

The pulsar current is used to derive constraints on the birth of pulsars from the statistical distribution, $N_P(P,{\dot P})$, of the population of pulsars in the $P$--${\dot P}$ plane (Figure~\ref{fig1}), where $P$ is the period and ${\dot P}$ is the period derivative.  Individual pulsars move along trajectories  on the $P$--${\dot P}$ plane, becoming observable (``birth'') at some point and disappearing from observation (``death'') at some other point. In a steady state there should be a constant flow of pulsars in the $P$--${\dot P}$ plane from the region where they are born to the region where they die. This flow is referred to as the pulsar current \citep{pb81,vn81,drs95}, which is defined as the integral over ${\dot P}$ of $J_P={\dot P}N_P(P,{\dot P})$. In a standard model \citep{lmt85}, all pulsars are thought to be born in supernovae with short initial periods. The expected pulsar current should increase as a function of $P$, reaching a plateau at $P=P_{0}$, with $P_0$ corresponding to the maximum period at birth, followed by decline at large $P$ as pulsars die. The birthrate is proportional to the height of the plateau. An early analysis using the pulsar current technique indicated a step-like increase in the pulsar current at around $P=0.5\rm\,s$ \citep{vn81}, implying significant injection at relatively large $P$, and also at relatively high magnetic fields ($B\geq 10^{12}\rm\,G$). However, \citet{lbdh93} found no strong evidence for injection at such relatively long periods. A more recent analysis \citep{vml+04} suggests that up to 40\% of all pulsars are born with periods in the range 0.1--$0.5\rm\,s$, and that over half the pulsars are born with magnetic fields $>2.5\times 10^{12}\rm\,G$. From the latest pulsar current analysis, \citet{kk08} found that the inferred birthrate is not obviously consistent with the supernova rate.  
 
In this paper we reconsider the formal basis for the pulsar current technique, questioning the view that it is ``model free.'' The formal basis for the pulsar current technique can be understood in terms of an analogy between the evolution of pulsars and the evolution of a one dimensional (1D) dynamical system \citep{pb81}. The period, $P$, plays the role of a generalized coordinate,  and the period derivative, $\dot P$, plays the role of a generalized velocity. The evolution of pulsars may be described by a force law, which can be written as ${\ddot P}=f(P,{\dot P})$, and the solution of this equation gives the trajectory (the path of an individual pulsar in the $P$--$\dot P$ plane), which depends on the specific function $f(P,{\dot P})$, and on an arbitrary constant of integration that can be different for different pulsars. Some examples were given by \citet{pb81}. Here we choose a widely favoured (``constant-$n$'') model, in which pulsars are assumed to evolve such that they have a constant braking index, $n$, to illustrate our arguments. 

The conventional interpretation of the pulsar current relies on two assumptions. One assumption is that all pulsars evolve according to the same law, and that this law depends only on $P$ and $\dot P$. The requirement that the evolution depend only on $P$ and $\dot P$ implies that it must not depend on any additional pulsar variable, such as a changing magnetic field, $B$, or a changing angle, $\alpha$, between the rotation and magnetic axes. This requires that pulsars with the same $P$ and $\dot P$ must have the same $\ddot P$ and the same $n$. This assumption seems inconsistent with glitching, involving discontinuous changes in $\dot P$, and other non-systematic changes observed in pulsar timing. The second assumption relates to the interpretation of $P$--$\dot P$ space as a phase space, which underlies the concept of a current in this space. This assumption is satisfied only for specific classes of force laws, which includes all ``potential'' force laws, in which the force $f(P,{\dot P})$ does not depend on $\dot P$. This assumption is violated for the constant-$n$ model. This inconsistency can be corrected by introducing a pseudo source term, that mimics births or deaths, but actually corresponds to convergence or divergence of trajectories, due to the wrong phase space being used. In this sense, the interpretation of the pulsar current is definitely not model free: for the constant-$n$ model one needs to include the pseudo source term that depends on $n$. One of our objectives in this paper is to show how this pseudo source term affects the interpretation of a specific set of pulsar data.

Despite these seemingly serious and perhaps fatal flaws, it is generally accepted that the pulsar current analysis leads to results for pulsar birthrates that are not inconsistent with other data, suggesting that the pulsar current approach is rather robust. It seems reasonable to assume that there is an average universal law for pulsar evolution, on which glitching and other random changes is superimposed. Assuming there is a universal evolution law, its form is not known. A constant-$n$ model is one of many different models that one can choose, and it has the unsatisfactory feature that it requires one to introduce the pseudo source term. There are other choices, including ``potential'' models, for which this complication does not occur. This leads to an obviously unsatisfactory situation in that the interpretation of the pulsar current seems to depend on an almost arbitrary assumption concerning the mathematical form of a putative evolution law. After demonstrating the complications that a constant-$n$ law introduces, we argue that it should be abandoned.  If there is an (average) evolution law for pulsars, the simplest assumption one can make about this law is that it has a potential form. 

In Section~\ref{sect:kinetic} the kinetic equation for the pulsar distribution is defined and discussed, and the pseudo source term is identified for a general class of evolution laws. In Section~\ref{sect:data} we show how the inclusion of the pseudo source term affects results on the birthrate derived from a statistical analysis of pulsar data. In Section~\ref{sect:alternative} we discuss existing mathematical models for the evolution, we extend the formal theory by introducing a Lagrangian formalism, and we carry out an analysis for a general form that includes both the constant-$n$ and potential laws as special cases. We discuss the results in Section~\ref{sect:discussion}.

\section{Kinetic equation for pulsars}
\label{sect:kinetic}

The distribution function for pulsars (number per unit area of the $P$--${\dot P}$ plane), $N_{P}(P,{\dot P},t)$, regarded as a function of the variables  $P$, ${\dot P}$, obeys a kinetic equation obtained by differentiating with respect to time. However, the resulting kinetic equation is not in the form (similar to Liouville's equation in statistical mechanics) that underlies the interpretation of the pulsar current. In this section we show that two forms of the kinetic equation differ by a pseudo source term.

\subsection{Evolution of the distribution of pulsars}

It was argued by \citet{pb81} that the motion of the observable pulsars in the $P$--${\dot P}$ plane (Figure~\ref{fig1}) can be interpreted as a ``fluid'' flow, where the fluid density corresponds to the density, $N_{P}(P,{\dot P},t)$, of pulsars in the $P$--${\dot P}$ plane. Specifically, $N_{P}(P,{\dot P},t)dP d{\dot P}$  pulsars with $P$ between $P$ and $P+dP$, and with ${\dot P}$ between ${\dot P}$ and ${\dot P}+d{\dot P}$. The rate of change of $N_{P}(P,{\dot P},t)$ with time is due to evolution of individual pulsars and to birth and death of pulsars. These are described by the three terms in
\begin{equation}
\frac{d}{dt}N_P(P,{\dot P})=
S_b(P,{\dot P})-S_d(P,{\dot P}),
\label{bmd}
\end{equation} 
where $S_b(P,{\dot P})$ and $S_d(P,{\dot P})$ are the rates, per unit time and per unit area of the $P$--${\dot P}$ plane, that pulsars are born and die, respectively.

According to the rules of partial differentiation, the time derivative of $N_P(P,{\dot P},t)$ on the left hand side of (\ref{bmd}) is given by
\begin{eqnarray} 
\frac{d}{dt}N_P(P,{\dot P},t) =  \frac{\partial N_P(P,{\dot P},t)}{\partial t} +  
  {\dot P} \frac{\partial N_P(P,{\dot P},t)}{\partial P}
\nonumber
\\
+{\ddot P} \frac{\partial N_P(P,{\dot P},t)}{\partial {\dot P}}.
\qquad\qquad
\label{kinetic1}
\end{eqnarray}
We are interested in a steady-state distribution of pulsars, $\partial N_P(P,{\dot P},t)/\partial t=0$, in which evolution and the birth and death rates are in balance. 

\subsection{Pulsar current}

The flow of pulsars through a vertical line in the $P$--${\dot P}$ plane is interpreted as the pulsar current. The pulsar current is conventionally identified based on the kinetic equation
\begin{equation}
\frac{\partial[ {\dot P}  N_P(P,{\dot P})]}{\partial P}
+ \frac{\partial[{\ddot P} N_P(P,{\dot P})]}{\partial {\dot P}}
=S_b(P,{\dot P})-S_d(P,{\dot P}).
\label{kinetic2}
\end{equation}
On integrating (\ref{kinetic2}) over $\dot P$, the integral over the term involving the ${\dot P}$-derivate is assumed to integrate to zero. The integral over ${\dot P}$ of 
\begin{equation}
J_P(P,{\dot P})={\dot P}  N_P(P,{\dot P}),
\label{current}
\end{equation}
is interpreted as the pulsar current. ($J_P(P,{\dot P})$ is often referred to as the pulsar current, but it is actually the current density in a localized range of ${\dot P}$.) The integral over ${\dot P}$ of
\begin{equation}
 \frac{\partial J_P(P,{\dot P})}{\partial P}=
 S_b(P,{\dot P})-S_d(P,{\dot P})
\label{kinetic3}
\end{equation}
then expresses conservation of pulsars. This allows one to infer properties of the source term, $S_b(P,{\dot P})$ describing birth of pulsars, and the sink term, $S_d(P,{\dot P})$ describing death of pulsars, from the statistical distribution of pulsars on the $P$--${\dot P}$ plane.

\subsection{Pseudo source term}

The kinetic equation (\ref{kinetic2}) differs from that obtained by combining  (\ref{kinetic1}) and (\ref {bmd}) in that the factors $\dot P$ and $\ddot P$ are inside the respective derivatives. To derive (\ref{kinetic2}) from (\ref{kinetic1}) one needs to move $\dot P$ and $\ddot P$ inside the respective derivatives. Assuming that $P$ and $\dot P$ are the independent variables, which is implicit in the use of the $P$--${\dot P}$ plane, the first of these is trivial. For the evolution law  ${\ddot P}=f(P,{\dot P})$ assumed by \citet{pb81}, the second introduces an extra term, which we move to the right hand side and interpret as a pseudo source term,
\begin{equation}
S_{ps}(P,{\dot P})= {\partial f(P,{\dot P})\over\partial{\dot P}}N_P(P,{\dot P}).
\label{pseudo}
\end{equation}
The right hand side of (\ref{kinetic3}) is then replaced by $ S_b(P,{\dot P})-S_d(P,{\dot P})-S_{ps}(P,{\dot P})$. In order to interpret information on the birthrate of pulsars from statistical data, one needs to calculate the additional term (\ref{pseudo}). We discuss the effect of this term on the interpretation of the pulsar current for evolution at fixed $n$ in the next section. We then argue that there is little evidence for the actual form of the pulsar evolution equation, that any model involves an ad hoc assumption, and that a sensible criterion to adopt in choosing an evolution equation is that the pseudo force term is absent. 

\section{Observational results}
\label{sect:data}

The effect of the pseudo source term on the interpretation of the pulsar current is explored in this section for the widely favoured model of evolution at fixed $n$. This corresponds to $f(P,{\dot P})={(2-n){\dot P}^2/P}$ in (\ref{pseudo}). The main qualitative point we make is that the use of the pulsar current is model-dependent, with the explicit dependence on the model appearing through the pseudo-source term.

\begin{figure}
\begin{center}
\includegraphics[width=70mm,height=80mm,angle=-90]{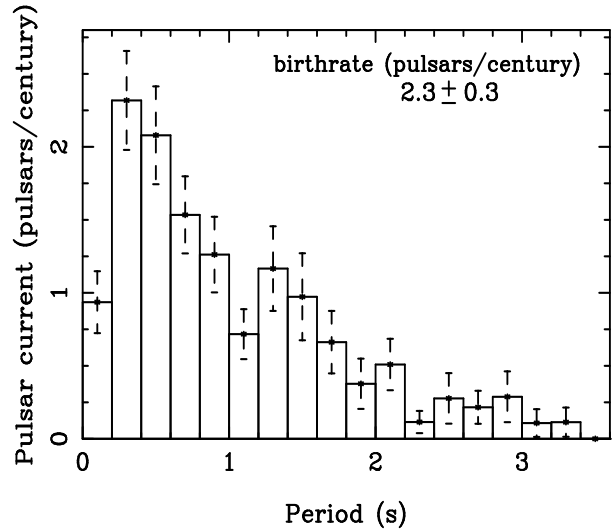}
\caption{\label{fig2}Updated standard pulsar current using 1438 non-recycled pulsars from Parkes surveys. The NE2001 Galactic electron density \citep{cl02} and the \citet{tm98} beaming models were adopted, for sources above a 1.4-GMz radio threshold of $0.1\,{\rm mJy}\,{\rm kpc}^{2}$. Under steady-state conditions, the maximum value of current is the birthrate of pulsars.}
\end{center}
\end{figure}

\begin{figure}
\begin{center}
\includegraphics[width=70mm,height=80mm,angle=-90]{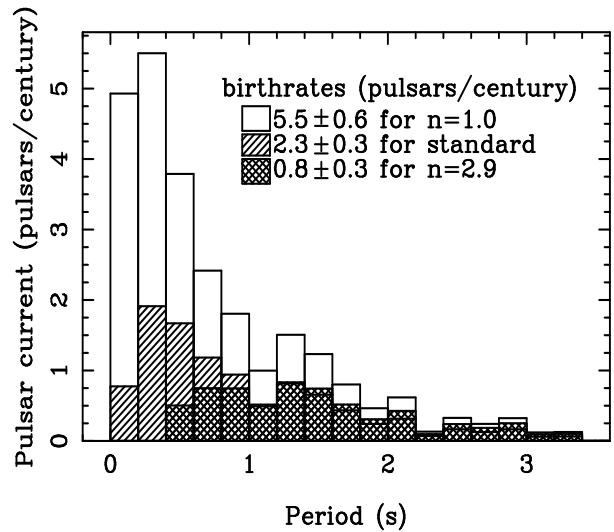}
\caption{Standard pulsar current (represented with hatched fill-histogram-area) versus revisited pulsar currents (represented with the outlined and cross-hatched fill-histogram-areas). Errors have been excluded for clear comparison of three current histograms. 
\label{fig3}}
\end{center}
\end{figure}

\subsection{Pseudo source term for different values of $n$}

We revisited our standard pulsar current analysis \citep{vml+04} by including the pseudo-source term in our calculation. In the following analysis we used 1438 non-recycled  pulsars that were predominantly discovered by the Parkes pulsar surveys, of which all are obtained from public pulsar catalogue at {\tt http://www.atnf.csiro.au/research/pulsar/psrcat}. %, with a mean period of ?0.76 s. 
We accurately modelled the sensitivity threshold of the multibeam survey and use this model to compute the pulsar birthrate.  
%1) placing it at large number of randomly selected locations in the model Galaxy; 
%(the pulsars are distributed uniformly in $\phi$ and exponentially with density 
%$\rho(R,z)$ in galactocentric radius R and height above the Galactic plane z) 
%2) calculating the effective dispersion measure and the interstellar scattering;
% using Taylor \& Cordes (1993) and Cordes \& Lazio (2002) electron density 
%models; 
%and 3) recording the number of detections, i.e. those positions for which the predicted flux density exceeds the survey limit. 
%\\S(P,L) is defined as ratio of the volume (weighted by pulsar density) of the
% Galaxy to the volume (also weighted by pulsar density) of the Galaxy where a 
%pulsar of period P and luminosity L would be detected. 
%For each pulsar, then, the scale factor is simply the ratio of the total number of locations in our model Galaxy (typically $10^5$) to the number of detections. This results in an estimate of the number of similar pulsars in the Galaxy. 
To simulate the uncertainties in the data, rather than choosing a fixed braking index, we chose $n$ randomly distributed about a mean, ranging from 1 to 5, and then calculated the pulsar current, with the same radio luminosity threshold, Galactic electron density and beaming models as for the conventional pulsar current approach \citep{vml+04}, Figure~\ref{fig2}. In Figure~\ref{fig3} we plot the standard pulsar current together with two revisited pulsar currents, denoting their corresponding pulsar birthrates. There are substantial differences in currents corresponding to different values in $n$. For $n>2$, the $S_{ps}(P,{\dot P})$ is negative, which reduces the pulsar current, implying that the neglect of the pseudo source term leads to an overestimate of the current. For $n<2$, $S_{ps}(P,{\dot P})$ is positive, implying that the neglect of the pseudo source term leads to an underestimate of the current.

\begin{figure}
\begin{center}
{\bf a)} 
\includegraphics[width=70mm,height=80mm,angle=-90]{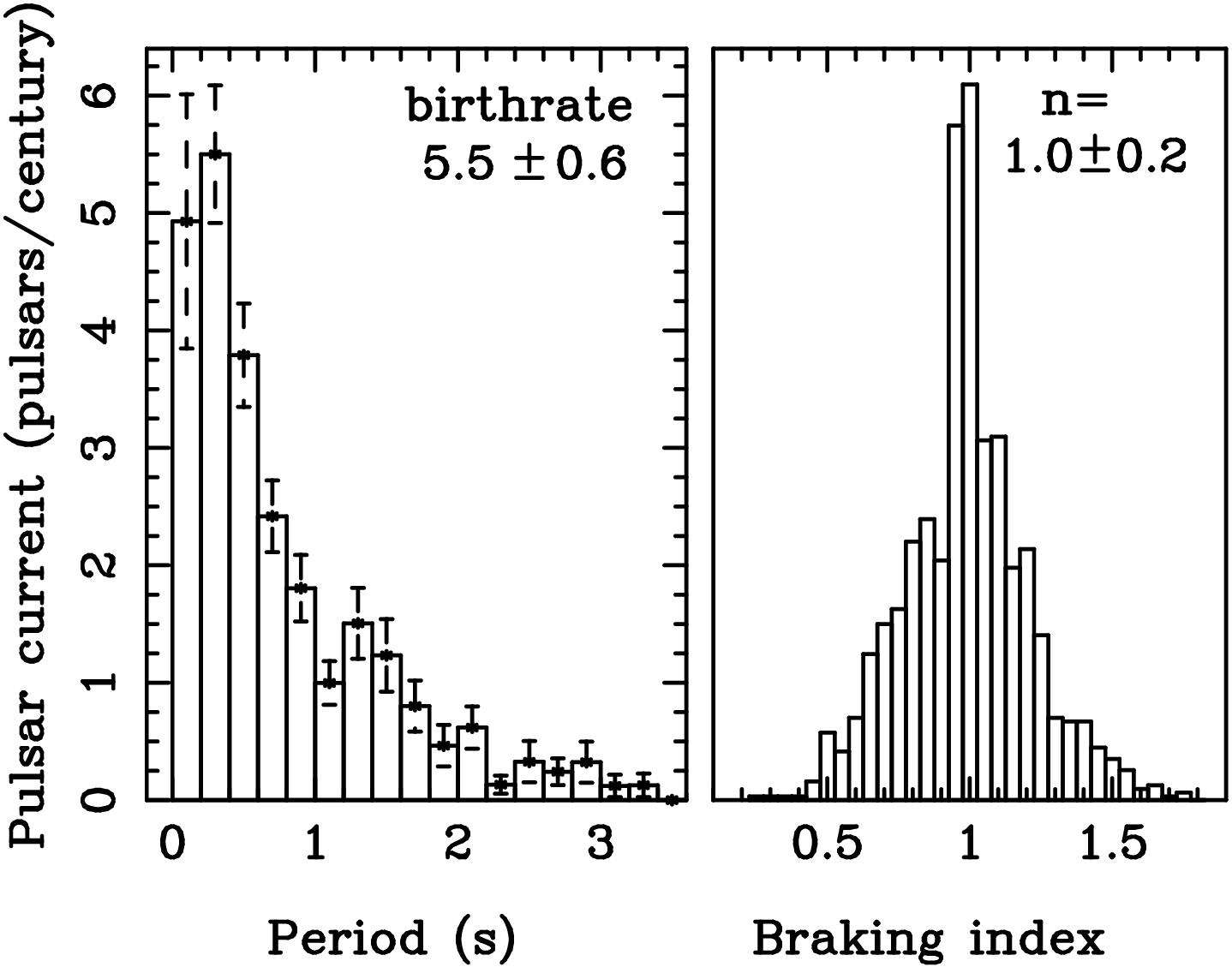} \\
{\bf b)}
\includegraphics[width=70mm,height=80mm,angle=-90]{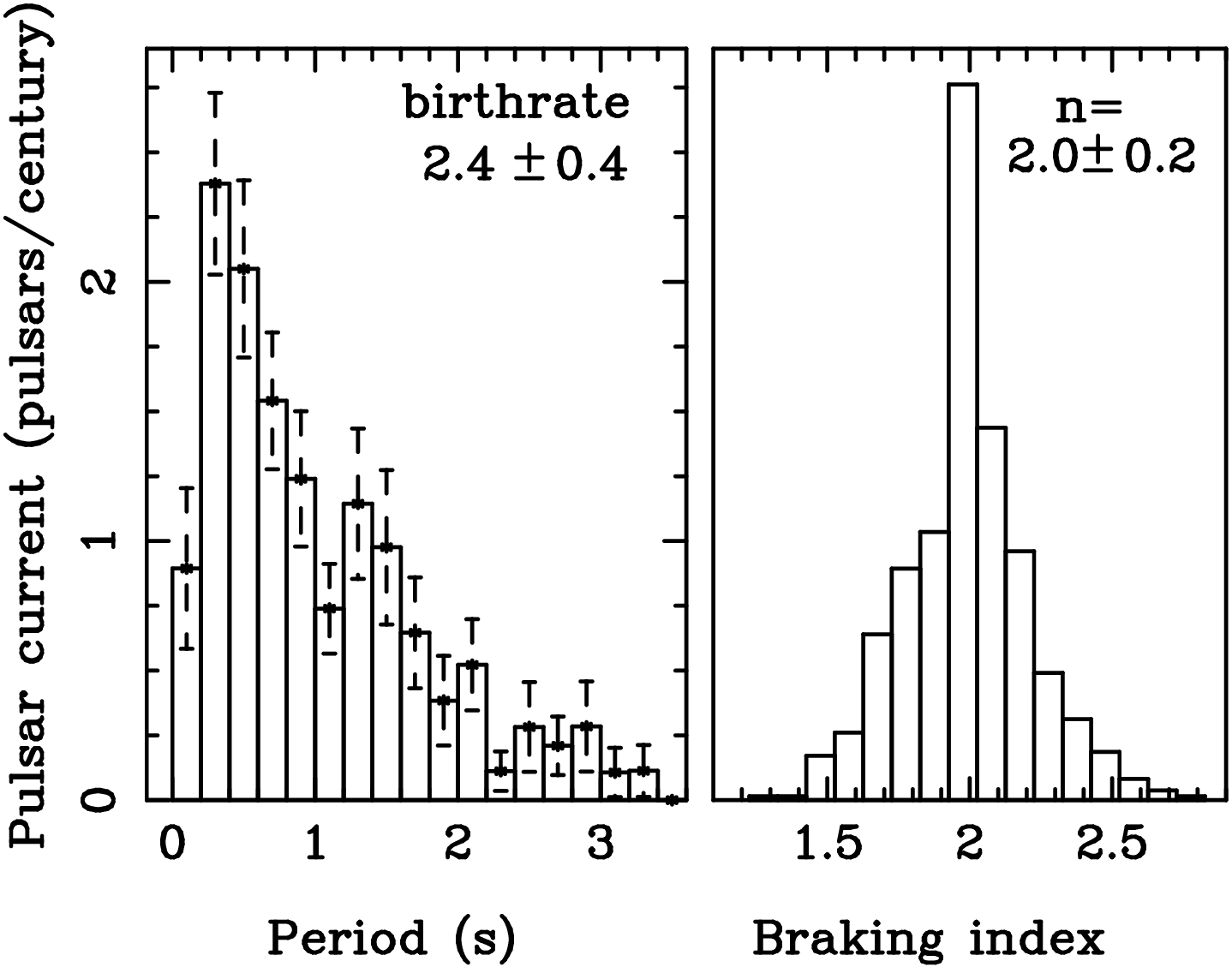} \\
{\bf c)}
\includegraphics[width=70mm,height=80mm,angle=-90]{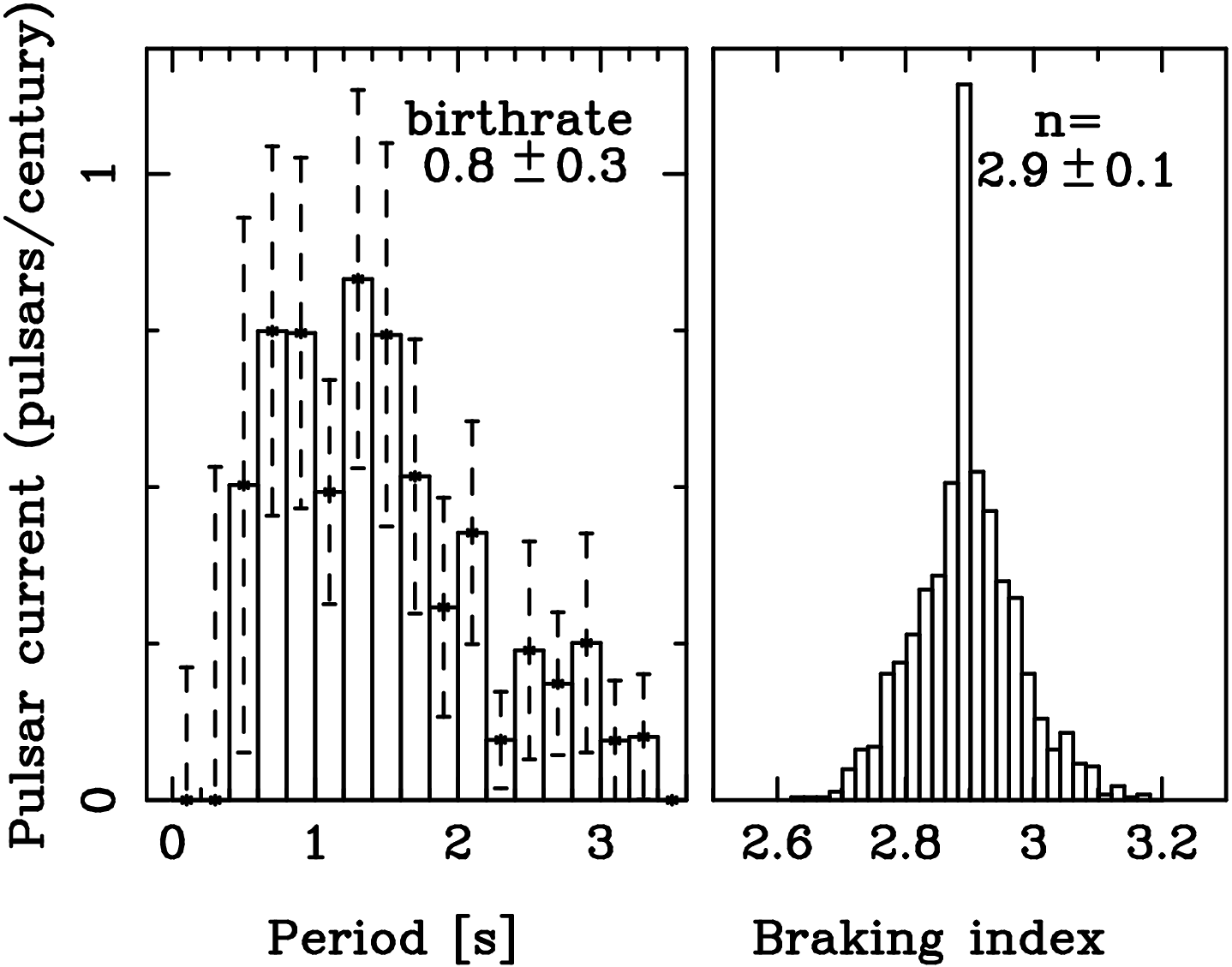}
\caption{Revisited pulsar current for different random braking indices. Each of a), b), and c) plots represents combination of pulsar current histogram on left (with birthrate written at the right top corner of the plot) and its corresponding histogram of randomly chosen braking indices on right (with the average braking index and its standard deviation written at the right top corner of the plot).}
\label{fig4}
\end{center}
\end{figure}

Our simulations for different runs of braking index are shown on Figures~\ref{fig4}~a), b) \& c). Each plot displays two histograms, the revisited pulsar current on the left, and its corresponding histogram of randomly chosen braking indices on the right. Matching  birthrate and average braking index with its standard deviation are displayed on each histogram. The effect of the assumed value of $n$ on the estimated birthrate for given data can be seen from  the plots: $n<2$ ($n>2$) the inferred birthrate is higher (lower) than when the pseudo source term is neglected.

\subsection{Measured braking indices}

\citet{lyn04} and \citet{lkg05} reported braking index measurements for six young pulsars (Crab, B0540-69, Vela, J1119-6127, B1509-58, and J1846-0258). Five have an unambiguous measurements of $n$, via phase-coherent timing, and these are marked by a star in Table~\ref{tab1}. We show their implied trajectories in the $P$--${\dot P}$ plane by arrows in Figure~\ref{fig1}. 

\begin{table*}
\begin{center}
\begin{minipage}{150mm}
\caption{Braking index measurements for six pulsars. Also given are the pulsar period, period derivative, period second derivative, and characteristic age. \label{tab1}}
%\end{minipage}
%\begin{minipage}{150mm}
\begin{tabular}{c c c c c c} \hline
\multicolumn{1}{c}{Pulsar names} & 
\multicolumn{1}{c}{$P$}  & 
\multicolumn{1}{c}{$\dot{P}$} &
\multicolumn{1}{c}{$\ddot{P}$} & 
\multicolumn{1}{c}{$\tau_{c}$} & 
\multicolumn{1}{c}{braking index} \\

\multicolumn{1}{c}{J2000; B1950} & 
\multicolumn{1}{c}{(ms)} & 
\multicolumn{1}{c}{$(10^{-15})$} & 
\multicolumn{1}{c}{($\rm\,s^{-1}$)} & 
\multicolumn{1}{c}{($\rm\,yr$)} &
\multicolumn{1}{c}{$\rm\,n$} \\ 
\hline 

Crab$^{*}$  & & & & & \\
J0534+2200; B0531+21 & 33.085 & 423 & $-3.61\times 10^{-24}$ & 1,240 & 2.51(1)\footnote{\citet{dp83, lps88, lps93}} \\
J0540-6919; B0540-69$^{*}$  & 50.499 & 479 & $-1.6\times 10^{-24}$ & 1,670 & 2.140(9)\footnote{\citet{lkg05}} \\
Vela & & & & &  \\
J0835-4510; B0833-45 & 89.328 & 125 & & 11,300 & 1.4(2)\footnote{\citet{lpgc96, dml02}} \\
J1119-6127$^{*}$ & 407.746 & 4022 & $-8.8\times 10^{-24}$ &1,160 &  2.91(5)\footnote{\citet{ckl+00}}  \\
J1513-5908; B1509-58$^{*}$  & 150.658 & 1540 & $-1.312\times 10^{-23}$ & 1,550 & 2.839(3)\footnote{\citet{mdn85, kms+94}} \\
J1846-0258$^{*}$  & 325.684 & 7083 & & 728 & 2.65(1)\footnote{\citet{lkgk06}} \\
\hline
\end{tabular}
\end{minipage}
\end{center}
\end{table*} 

There are only a few pulsars that are potential candidates for accurate measurement of $n$, which requires that they satisfy the following. Firstly they must spin down sufficiently quickly to allow a useful measurement of $\ddot{P}$. Secondly, the position of the pulsar should be accurately known at the $\sim$1 arcsecond level. Thirdly, the spin-down must not be seriously affected by glitches, sudden spin-ups of the pulsar, or timing noise. Typically, glitches begin to seriously affect smooth spin down at characteristic ages of $\sim 5-10$\,kyr \citep{ml90,mgm+04}. Thus many of the pulsars that may spin down fast enough for a measurement of $n$ are irretrievably contaminated by glitches \citep{sl96,wmz+01}. Timing noise varies from object to object,  is roughly correlated with spin-down rate \citep{antt94} and can prevent a measurement of $n$ in a finite data set in an unpredictable way. All six trajectories have $n<3$,  three of them have upward slopes ($n<2$), and none of them is directed towards the main body of pulsars. These data do not support the constant-$n$ model, which requires the same value of $n$ for all pulsars at all $P$, $\dot P$.

\section{Alternative evolution laws}
\label{sect:alternative}

For the pulsar current to be a well-defined concept, there must be a universal law of the form ${\ddot P}=f(P,{\dot P})$ obeyed by all pulsars. This is clearly not the case. For the pulsar current to be a useful concept, such an evolution law must apply at least in some average sense for all pulsars. The constant-$n$ evolution law assumed in the previous section, although widely favoured, is essentially ad hoc. In this section we introduce a general (Lagrangian) formalism, that assumes that there is a unique evolution law but makes no assumptions on its form. We use this formalism to identify a simpler class of evolution models that we refer to as potential models.

\subsection{Specific models}

Before introducing a general theory for a possible pulsar evolution model, we summarize the models that underlie the discussion in the previous section.

\subsubsection{Vacuum dipole model}

The vacuum dipole model has a contradictory status in pulsar physics. One the one hand, it is  the basis of our interpretation of pulsars, notably $B\sin\alpha\propto(P{\dot P})^{1/2}$, and it is the only model for which the evolution is well determined. On the other hand, the model is clearly inapplicable to actual pulsars, and the actual evolution is inconsistent with observations. The constant $n=3$ model is often assumed to describe evolution in the vacuum dipole, but this is incorrect. 

The model is that a (structureless, point) rotating magnetic dipole in vacuo radiates electromagnetic radiation that escapes to infinity, carrying off energy and angular momentum. The loss of energy causes $P$ to increase on a characteristic spin-down timescale, and the torque causes $\sin\alpha$ to decrease, implying that the rotation and magnetic axes tend to align on the spin-down time scale \citet{mic91}. The braking index is $n=3+2\cot^2\alpha$ \citep{mic91}, which is always greater than 3, and which increases as $\sin\alpha$ decreases. 

This vacuum-dipole model, at least in its simplest form, is unacceptable for pulsars. The putative electromagnetic radiation has a frequency equal to the rotation frequency of the pulsar, well below the plasma frequency in the pulsar wind and in the ISM, and it cannot escape to infinity, as the model requires. The model predicts that the magnetic and rotation axes align on the spin-down time scale. There is a theoretical argument that the structure of the star may slow the alignment \citep{g70}, and there is evidence for older pulsars having smaller of values of $\sin\alpha$ \citep{wj08}, but the prediction of the model itself  is inconsistent with pulsar data.

\subsubsection{The $n=3$ model}

The constant $n=3$ model is often confused with the vacuum dipole model. In the $n=3$ model, it is assumed that the evolution of pulsars is at constant $B\sin\alpha$, and this is combined with the vacuum dipole model so that the assumption becomes evolution at constant $P{\dot P}$. Note that this is inconsistent with the actual vacuum dipole model, which implies systematically decreasing $\sin\alpha$. Thus, the basic assumption in the constant $n=3$ model is that the trajectory of pulsars in the $P$--${\dot P}$ is of the form ${\dot P}=K/P$, where $K$ is a constant.  The force law implied by this relation is ${\ddot P}=-{\dot P}^2/P$ and this implies a braking index
\begin{equation}
n=2-{{\ddot P}P/{\dot P}^2}
\label{braking}
\end{equation}
equal to 3. Although a vacuum model modified by assuming $B\sin\alpha$ remains constant is widely favoured it is not supported by any detailed theory for the evolution.

\subsubsection{$n=$ constant model}

Another model that is used widely is based on the assumption that the braking index is constant but not necessarily equal to 3; we refer to this as the $n=$ constant model. According to (\ref{braking}), $n=$ constant implies ${\ddot P}=(2-n){\dot P}^2/P$, and integrating this equation leads to the trajectory  ${\dot P}=KP^{2-n}$, where $K$ is a constant. This is the model used for illustrative purposes in the previous section. However, the assumption $n=$ constant on which it is based is also not supported by any detailed theory for the evolution, and should be regarded as ad hoc.

\subsubsection{Potential models}

Granted that one needs to make an ad hoc assumption in choosing a possible universal law governing pulsar evolution in  $P$--${\dot P}$, it is desirable to choose a law that does not introduce unnecessary complications. As shown in detail above, the $n=$ constant model leads to the need to include a pseudo source term in the interpretation of the pulsar current. A model that avoids this complication is one in which $\ddot P$ depends only on $P$. We refer to this as a potential model. The law is written in the form $\ddot P=-C'(P)$, where $C(P)$ is the potential in the theory developed below. This theory is also not supported by any detailed theory for the evolution, and is also ad hoc. However, a potential model has the advantage of being the simplest model, and the best justification for it is that one should explored the simplest model first.

\subsection{Lagrangian formulation}

In a Lagrangian formulation of dynamics the independent variables are the generalized coordinates and generalized velocities. In the present case, the generalized coordinate is identified as $P$ and the generalized velocity as $\dot P$. A Lagrangian, $L(P,{\dot P})$ may be identified by requiring that the Euler-Lagrange equation
\begin{equation}
{d\over dt}{\partial L(P,{\dot P})\over\partial{\dot P}}-{\partial L(P,{\dot P})\over\partial P}=0,
\label{euler}
\end{equation}
reproduce the equation of motion. 

Consider the evolution law \citep{pb81}
\begin{equation}
{\ddot P}={(2-n){\dot P}^2\over P}-{{\dot P}\over\tau}.
\label{PB1}
\end{equation}
This has a Lagrangian counterpart for $\tau\to\infty$. The relevant Lagrangian is a particular case of
\begin{equation}
L(P,{\dot P})={\textstyle{1\over2}}A(P){\dot P}^2+B(P){\dot P}-C(P),
\label{Lagrangian1}
\end{equation}
for which (\ref{euler}) implies
\begin{equation}
{\ddot P}=-{A'(P)\over2A(P)}{\dot P}^2-{C'(P)\over A(P)},
\label{Lagrangian2}
\end{equation}
where the prime denotes a derivative in $P$. The form (\ref{PB1}) is reproduced for $A(P)=P^{-2(2-n)}$, $C(P)=0$. 

The trajectory of a pulsar in $P$--${\dot P}$ space is found by solving the equation of motion. This is facilitated by noting the analogy with a one-dimensional particle in a potential, with $A(P)$ playing the role of the mass and $C(P)$ playing the role of the potential energy. The energy-integral (sum of kinetic and potential energies equals a constant) gives
\begin{equation}
{\textstyle{1\over2}}A(P){\dot P}^2+C(P)=E,
\label{Lagrangian3}
\end{equation}
where $E$ is a constant of integration (the counterpart of the energy). A second integral gives
\begin{equation}
\int dP\,\left({A(P)\over E-C(P)}\right)^{1/2}=2^{1/2}t.
\label{Lagrangian4}
\end{equation}
The example considered by \citet{pb81} is reproduced by writing $E={1\over2}v^2$, $x=vt$.

\subsection{Potential models}

There is no pseudo source term in the kinetic equation for a Lagrangian of the form (\ref{Lagrangian1}) with $A(P)=1$. The generalized momentum, $\partial L/\partial{\dot P}$, is then equal to the generalized velocity, so that the $P$--$\dot P$ is a phase space in the sense required for Liouville's theorem to apply. In this case (\ref{Lagrangian2}) gives
\begin{equation}
{\ddot P}=-C'(P),
\label{Lagrangian5}
\end{equation}
and (\ref{Lagrangian3}), (\ref{Lagrangian4}) apply with $A(P)=1$. The braking index in this model is
\begin{equation}
n=2-{{\ddot P}P\over{\dot P}^2}=2+{PC'(P)\over2[E-C(P)]}.
\label{Lagrangian6}
\end{equation}
A simple limiting case is $E\gg C(P)$, when the evolution law reduces to ${\dot P}\approx$ constant, with a braking index $n\approx2$. 

In these two examples of evolution laws, the trajectories of individual pulsars depend on a constant of integration, which depends on the particular pulsar. For trajectories of the form ${\dot P}=KP^{2-n}$ this constant is $K$, and for the potential law (\ref{Lagrangian3}), the constant is $E$. For any given law, the constant of integration is determined for a particular pulsar by the values of $P$ and ${\dot P}$ at birth, and the subsequent evolution is determined by (\ref{Lagrangian3}) with $E$ fixed, but different for different pulsars.

\section{Discussion and conclusions}
\label{sect:discussion}

The theory for the pulsar current is based on the assumption that there is a universal evolution law for pulsars, and that it depends only on $P$ and $\dot P$. This requirement is not satisfied, with glitching and other random changes implying that the trajectory of at least some pulsars on the $P$--$\dot P$ plane is influenced by other effects. Nevertheless, it is possible that the evolution of all pulsars does follow a universal law in some average sense, and this would suffice to justify the pulsar current technique. Assuming there is an average universal evolution law that applies to all pulsar, it is clear that we have little information on what this law is. A constant-$n$ law has been used widely in the literature. We assume this law in Section~\ref{sect:data}, argue that such a law invalidates a conventional pulsar current analysis, and show how the error can be corrected by introducing a pseudo source term. We argue that there is no theoretical basis for a constant-$n$ law, and that the complication it introduces through pseudo source term makes it an undesirable choice for the form of the evolution law. We argue that the choice of a potential law for the evolution avoids the complication of a model-dependent pseudo source term.

We discuss evolution laws in Section~\ref{sect:alternative}. We comment on the misconception that a constant $n=3$ law corresponds to the vacuum dipole model, whereas it actually corresponds to a hybrid model, with $B\sin\alpha\propto(P{\dot P})^{1/2}$ determined by the energy loss rate in the vacuum dipole model, and with $B\sin\alpha$ assumed constant. The actual vacuum dipole model implies that $\sin\alpha$ decreases (alignment of the magnetic and rotational axes) on the spin-down time scale, and this does not occur. The vacuum dipole model is not applicable to actual pulsars, for this and other reasons. 
A general class of evolution models is developed in Section~\ref{sect:alternative}; this class includes the constant $n$ model, with $\ddot P=(2-n){\dot P}^2/P$, and potential models, with ${\ddot P}=-C'(P)$ where $C(P)$ is the potential. The evolution law is integrated to find the trajectory of the pulsar in $P$--$\dot P$ space. The trajectory can be written ${\dot P}=KP^{2-n}$ for a constant-$n$ model, and as ${\dot P}=\sqrt{2}[E-C(P)]^{1/2}$ for a potential model. The trajectory depends on a constant of integration, which is different for different pulsars. A physical model for the evolution is required to give a physical interpretation to such constants of integration. In the constant $n=3$ model, the constant $K$ is proportional to $B\sin\alpha$, but there is no obvious interpretation of $K$ for other models with constant $n\ne3$. A physical basis for a specific potential model is required to give a physical interpretation to $E$. 
 
Assuming a potential law opens up a new way of thinking about the evolution of pulsars in the $P$--$\dot P$ plane. In our formulation the potential, $C(P)$, is an arbitrary function of $P$. By exploring different choices of $C(P)$ one can ask what choice (if any) best fits the data. Although we do not attempt to do this here, we speculate briefly on one implication. Suppose that the ``potential energy'' is much smaller than the ``kinetic energy'' $|C(P)|\ll{1\over2}{\dot P}^2$. Then all trajectories have ${\dot P}\approx$ constant, implying a braking index $n\approx2$. The measured values of $n$ include $n<2$ and $n>2$, which could be interpreted either according to (\ref{Lagrangian6}), requiring $C'(P)<0$ and $C'(P)>0$, respectively, or interpreted as random variations about an average universal $n\approx2$ for young pulsars. Perhaps average braking indices can be estimated for a large number of pulsars over a wide range of $P$, as suggested by \citet{jg99} for example. If such an estimation of $n$ shows some systematic trend with $P$, (\ref{Lagrangian6}) would provide some constraint on the possible form for $C(P)$. With such data one should be able to draw conclusions on whether simple examples potential law are consistent with observations; simple models include $C(P)\propto P^a$ with $a>0$, and $C(P)\propto -P^{-b}$ with $b>0$.  If there is no systematic trend, even in an average sense, this would cast serious doubt on the validity of the pulsar current technique, and conclusions drawn using it. 

In conclusion, the pulsar current approach is not ``model free'' and its interpretation depends on assumptions made about the form of the evolution law for pulsars. The theory depends on this law being universal, and although this is clearly not the case in an exact sense, it may be the case in some average sense. In modelling such an average law, the constant-$n$ assumption should be abandoned, and a ``potential'' law should be the preferred choice on which to base future discussions. 

\section*{Acknowledgments} 
The Parkes telescope is part of the Australia Telescope which is funded by the Commonwealth of Australia for operation as a National Facility managed by CSIRO. We thank Dick Manchester for advice and informative comments. D. B. M. thanks Mike Wheatland for helpful discussions on the kinetic equation.

\end{document}